# Improvement in the Performance and Efficiency on Self-Deficient CaTiO$_3$: Towards Sustainable and Affordable New-Generation Solar Cells

Shashi Pandey[1], Alok Shukla[2*], Anurag Tripathi[1]

[1]Department of Electrical Engineering IET Lucknow, Uttar Pradesh 226021, India
[2]Department of Physics, Indian Institute of Technology Bombay, Powai, Mumbai 400076, India
Email: 2512@ietlucknow.ac.in, shukla@phy.iitb.ac.in*, anurag.tripathi@ietlucknow.ac.in

**Abstract:**

Thin films of pure and self-deficient calcium titanites i.e., CaTiO$_3$, Ca$_{1-\alpha}$Ti$_1$O$_3$, CaTi$_{1-\beta}$O$_3$ and CaTiO$_{3-\gamma}$ have been deposited on ITO substrate using dip coating method. X-ray diffraction and Scanning electron Microscopy (SEM) analysis confirm the structural and morphology of all deposited thin films. Photocurrent measurement has been done, and it is observed that during the incidence of UV light on the as-prepared device (i.e., in the "ON" state), a significant increase in photocurrent ($I_{UV}$) at zero voltage was observed in case of O deficient CaTiO$_3$, while in case of Ti and Ca deficient thin films smaller values of photocurrents were seen. Responsivity and detectivity of deposited thin films of all self-deficient CaTiO$_3$ were found to be maximum in the UV region while they also showed smaller contributions in the visible range possibly due to the presence of self-deficiency. The self-deficient sample exhibits lower resistance (higher recombination rate) than the pure sample at low voltage, but at higher voltage, it is almost identical. Furthermore, theoretical calculations have been performed using the first-principles density-functional theory to validate the experimental findings on self-deficient CaTiO$_3$. Incident-photon-to-current efficiency (IPCE) and current density have also been measured for all deficient CaTiO$_3$ samples. Maximum IPCE have been found in the range 25%-28% for Ti and O deficient samples in the UV range (280-400 nm). We argue that first-principles DFT calculations combined with the experimental measurements on self-deficient CaTiO$_3$ thin films offer a reliable way to enhance the performance of perovskite-based solar devices.

**Keywords:** *X-ray diffraction, Thin Films, Photocurrent, Detectivity, self-deficient, IPCE*

1.  **Introduction**

Semiconducting oxides, specially perovskites, are probably one of the most promising new energy options for solar cell based industry out of the many available ones[1–3]. Solar cell based devices use photovoltaic effects or photochemical reactions to transform energy of photons directly into electrical energy[4,5]. Calcium titanate ($CaTiO_3$) with the molecular structure of the type $ABO_3$, is a member of the perovskite family [6,7]. Perovskite materials have attracted a lot of attention these days because of their octahedral nanostructures and recent developments in synthesizing layered heterostructures. The lack of device stability of pure perovskite-based cells is now holding them back and limiting their usefulness[6,8,9]. Studies on the stability and rate of performance of the purest form of perovskite-based solar cells reveal that such devices are less stable and have lower rates of performance, as a result of which their commercial applications are limited. Recent research in the field of perovskite solar cells (PSCs) has shown that vacancy, chemical composition, temperature, UV radiation, and solution processing methods all have significant impacts on the structural stability and performance of solar devices[10–14]. It is well known that in comparison to their pure form, vacancies in perovskites [12,15] lead to enhancement in their electrical, opto-electronic and magnetic properties. Due to the presence of defects and the requirement of charge neutrality, one observes ionic transport in semiconductors because of the hopping of charges [16,17] at different sites of crystal. As a result, their electrical, magnetic and opto-electronic properties show significant changes when compared to the pure phase. The photovoltaic research community has been intrigued by the emergence of perovskites as solar cell materials because of how effectively they convert sunlight into energy. Their power conversion efficiency has increased from their initial debut efficiency of 3.8% in 2009 to the recently measured 25.2%[18].

For a material, the presence of optically active states in the visible region has several potential applications in photocatalysis, solar cells[19,20], storage devices, etc. Therefore, it is of utmost importance to investigate how vacancies impact the optical characteristics of $CaTiO_3$, the material which is the object of this study.[21,22]. The field of transition metal oxides (TMOs) based electronics [23] and related superlattices has been steadily expanding, and calcium titanate (CTO) has played a significant part in both. In TMOs, oxygen vacancies ($O_v$) are a common type of point defect[10,24–26], and their presence significantly impacts their electrical, opto-electronic and mechanical properties. In light of this, we conducted a theoretical and experimental analysis to examine the effects of the defect caused by self-doping in $CaTiO_3$.

In this work, we have performed a systematic experimental study of self-doped $Ca_{1-\alpha}TiO_3$, $CaTi_{1-\beta}O_3$ and $CaTiO_{3-\gamma}$ to investigate their optoelectronic properties. The experimental work has been supported by first-principles density-functional theory (DFT)-based theoretical calculations with the aim of understanding the enhancement in the stability and performance of the CTO devices with native defects. For the estimation of solar cell efficiency, IPCE and current density have also been measured for all deficient $CaTiO_3$. The present study suggests that by combining DFT calculations with the experimental measurements on deficient $CaTiO_3$ thin films, one can proceed in a systematic way to improve the stability and efficiency of perovskite-based solar cell devices.

## 2. Experimental and Theoretical Details:

### 2.1 Sample Preparation and Experimental Details:

Ethanol was mixed in $CaCO_3$ and $TiO_2$ powder with 1:7 molar ratio, and the mixture was stirred for two hours at the room temperature. The slurries were then dried for one hour in an oven set to 100°C. To create fine $CaTiO_3$ powder, the combined powders were crushed and then sintered in an air furnace for two hours at 900°C with a ramp rate of 3°C/min. The same procedure was used to make fine powder of self-deficient $CaTiO_3$ with the appropriate adjustment in molar fractions. The $ITO/TiO_2/CaTiO_3$ mesoporous heterojunction arrangement was used to construct PSCs devices based on pure and self-deficient $CaTiO_3$. A mild detergent, distilled water, and ethanol were used in order to clean the ITO glass in the ultrasonic bath. Additionally, $CaTiO_3$ was dissolved in ethanol and thoroughly mixed until dissolved at room temperature to create the $CaTiO_3$ paste. $TiO_2$ and $CaTiO_3$ paste were progressively applied to the cleaned ITO substrate using the dip-coating method. Carbon was then coated to serve as a counter electrode. On the PSCs' surface, in the area where they were active, KI and $I_2$ solution were dropped. The active area of the device was 10 $cm^2$. Zeiss Supra-55 field emission scanning electron microscope (FESEM) was employed for morphological study. The crystallinity of the sample was investigated using XRD using a Cu-Kα source. Using a UV-Vis spectrophotometer, absorbance spectra were examined in the wavelength range from 300-800 nm. Keithley 2612A source meter was used to accomplish the electrical characterizations.

## 2.2 Computational Details:

The plane-wave density-functional theory (PW-DFT) as implemented in the Vienna ab-initio simulation package (VASP) [12–14] was used to do the first-principles calculations to support the experimental findings. The generalised gradient approximation paired with the Hubbard U (GGA+U) [10,11,24,25] level of theory was employed to align our computed bandgap with the experimental findings. When the system's total energy is steady within $10^{-3}$ mRy, the self-consistent computations were deemed to have converged[26,27]. Using complicated dielectric functions and the Ehrenreich and Cohen formalism[28], the optical characteristics (absorption coefficient) of $CaTiO_3$ with and without point defects were obtained [24].

### 2.2.1 Calculations of Formation Energy:

The formation energies of the vacancies $V^\alpha$ ($\alpha$ = Ti, Ca, O) were calculated using the supercell approach. Here we used a 3 x 2 x 2 supercell for self-deficient calculations.

For the defect $D$ in charge state $c$, the formation energy $\Delta H(D^c)$ is calculated as

$$\Delta H(D^c) = E_{tot}(D^c) - E_{tot}(bulk) - \sum \Delta n_j \mu_j + c(E_v + E_F + \Delta V), \qquad (1)$$

where $E_{tot}$(bulk) and $E_{tot}(D^c)$ are, respectively, the total energies of the pure and self-deficient supercells of $CaTiO_3$, and $E_F$ is the Fermi energy of the system measured from the VBM. $\Delta n_j$ is the number of atoms of species $j$ ($j$ = Ca, Ti and O) that are removed from the supercell for creating defects. Finally, $\mu_j$ represents the chemical potential for the species $j$.

**Table1:** Formation energies of the pristine as well as self-deficient $CaTiO_3$

| Sample | $CaTiO_3$ | $Ca_{1-\alpha}TiO_3$ | $CaTi_{1-\beta}O_3$ | $CaTiO_{3-\gamma}$ |
|---|---|---|---|---|
| Formation energy | -0.5756 | 0.01327 | 0.02529 | 0.012932 |

From Table-1 it is obvious that the formation energy for creating oxygen vacancies is the minimum indicating that it is the most important and inherent defect in perovskite oxides.[10,24].

## 3. Results and Discussions:

### 3.1 X-ray diffraction and SEM analysis:

Figure 1 shows X-ray diffraction pattern of prepared self-deficient CTO samples after heat treatment at 900 °C. Prepared samples exhibit highly pure crystalline phase of $CaTiO_3$ with the growth direction along the (1 2 1) plane. Diffraction peaks are well matched with JCPDS No. 88-0790 [8,29] corresponding to an orthorhombic structure (see Figure 1), with the pbnm space group, and lattice parameters a = 5.441 Å, b = 7.643 Å and c = 5.381 Å. It suggests that using the sol-gel dip coating process and the aforementioned molar ratio, high crystalline pure and self-deficient CTO samples can be prepared. Figures 2 (a)-(d) show the morphological studies of pristine as well as self-deficient $CaTiO_3$ based deposited solar devices. It is clearly observed from SEM that the deposition is in the range of 100-150 nm of deficient CTO on ITO substrate. SEM images also indicate that the deposition has been done homogeneously on ITO substrate.

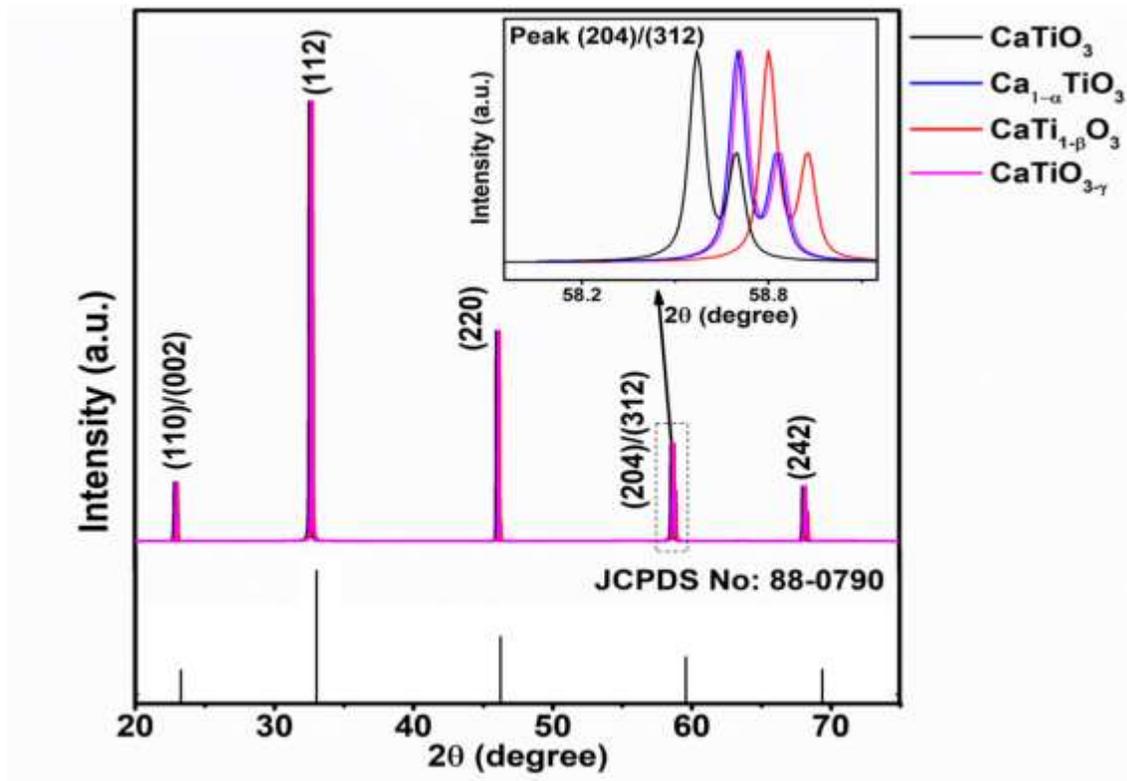

*Figure 1: X-ray diffraction pattern of pristine and self- deficient $CaTiO_3$ samples and diffraction peaks are well matched with JCPDS No. 88-0790.*

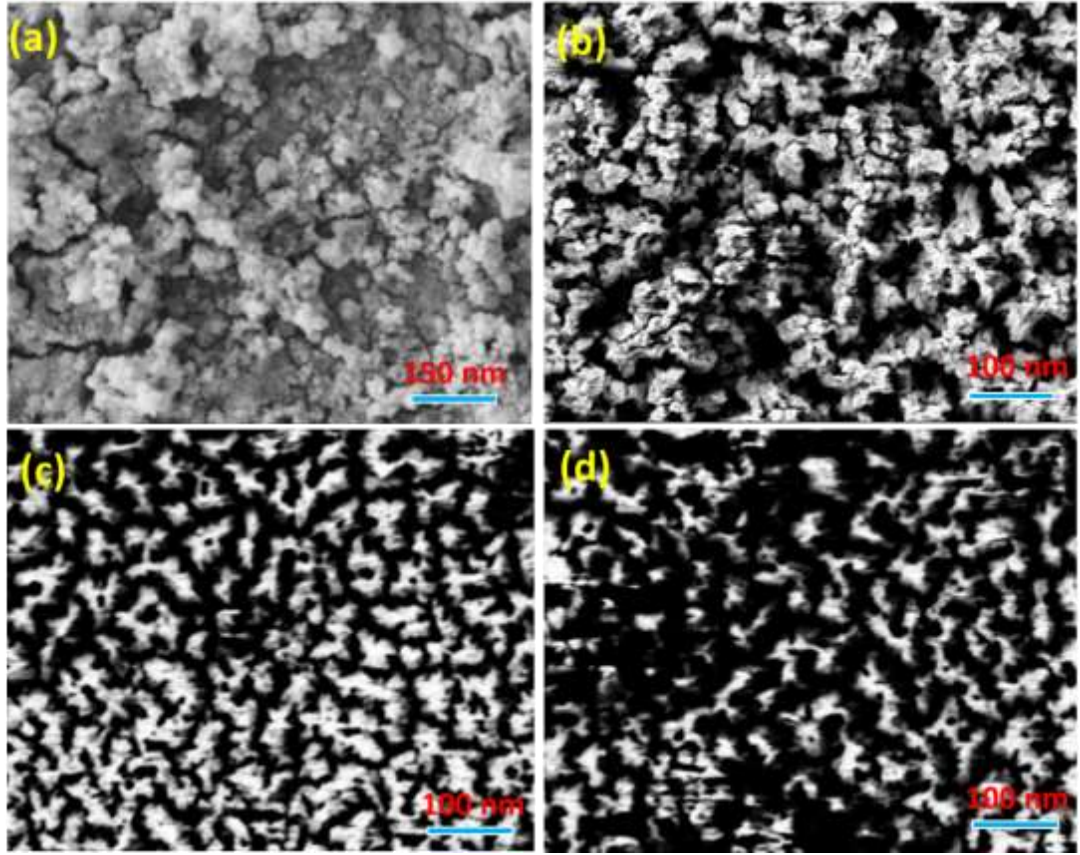

*Figure 2:* *Scanning Electron Microscope images of (a) CaTiO$_3$ (b) Ca$_{1-\alpha}$TiO$_3$, (c) CaTi$_{1-\beta}$O$_3$ and (d) CaTiO$_{3-\gamma}$.*

### 3.2. Bandgap Calculations using Density Functional Theory:

Total density of states (TDOS) computed using the DFT for all types of self-deficient CTO structures are shown in Figure 3. The calculations were performed using 3x2x2 super cells for the pure CaTiO$_3$ as well as those with vacancies at the Ca, Ti and O sites. The GGA+U exchange correlation functional was considered for all calculations [25,30]. From Figure 3 it is clear that an additional peak now been appears in between the conduction and the valance bands in TDOS [30,31] which is absent in pure CaTiO$_3$. It should be noted that the energy position of new absorption features is governed by type of vacancies present in the sample. In case of self-deficiency, states near fermi level are controlled by the O-2p and Ti-3d orbitals, signifying rehybridization between the orbitals of Ti-3d: e$_g$ and t$_{2g}$ with O-2p orbitals (i.e. p$_x$, p$_y$, p$_z$).

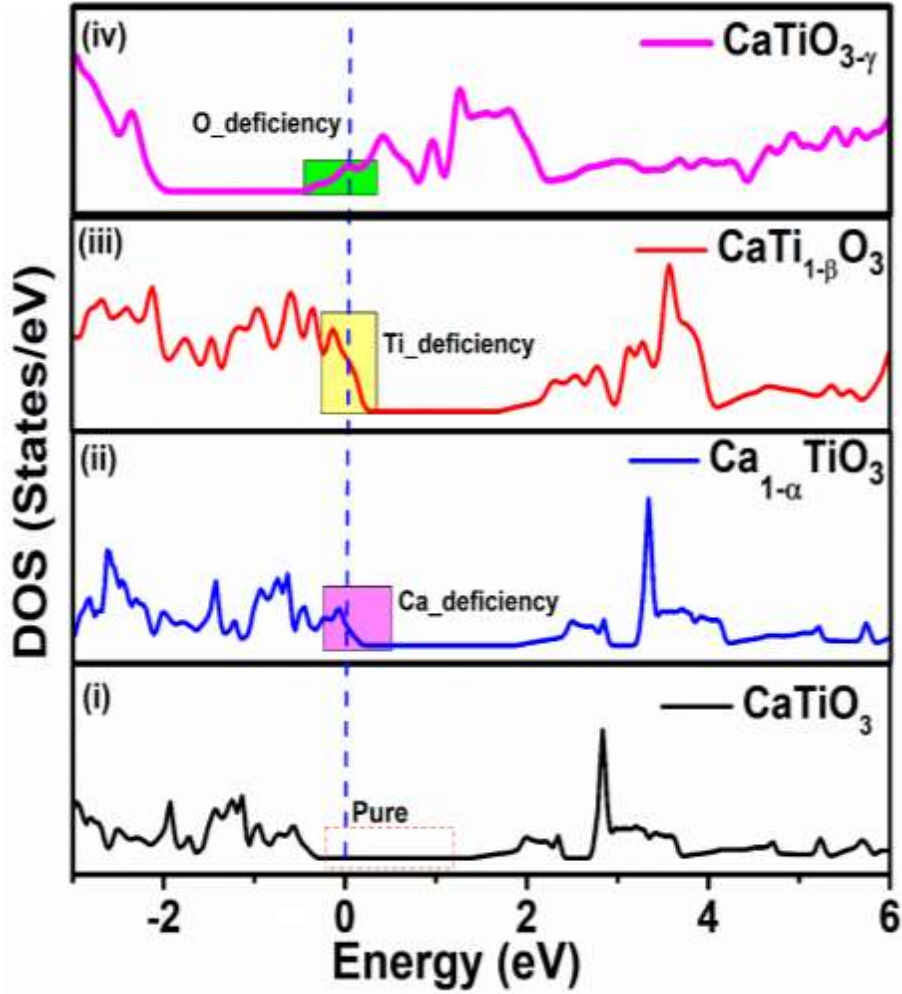

*Figure 3:* The total d*ensity of states of $CaTiO_3$ for: (i) pristine, (ii) Ca-deficient (iii) Ti-deficient, and (iv) O-deficient supercells.*

We have used the diffuse reflectance spectroscopy (DRS) to obtain the optical absorption for deficient $CaTiO_3$ samples. Observed spectra from DRS have been converted into equivalent absorption coefficient using Kubelka–Munk equation[32,33].

$$F(R_\infty) = \frac{(1-R_\infty)^2}{2R_\infty}, \tag{I}$$

shows Kubelka–Munk function in equation (I). To calculate the bandgap ($E_g$), the measured absorption coefficient is fitted to the Tauc equation[11] and plotted in figure 4. Tauc equation is determined as follows-

$$(\alpha h\nu)^n = A(h\nu - E_g), \tag{II}$$

In equation (II) above we use n=1/2 [34]for the deficient CaTiO$_3$ samples (for the indirect band-gap). Figure 4 shows the experimental band gap of deficient CaTiO$_3$ samples and lower energy transitions in optical absorption spectra have also been clearly observed.

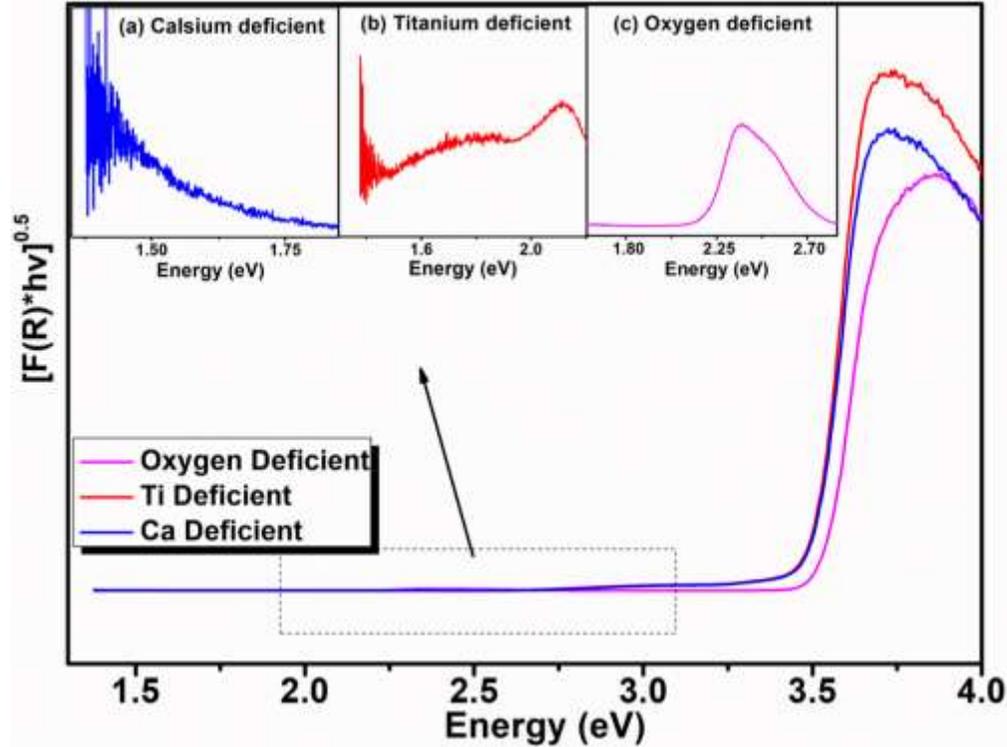

**Figure 4:** *Diffuse reflectance spectra of deficient CaTiO$_3$ samples.*

To compare our experimental with theoretical calculations, we have computed optical spectra for the self-doped CaTiO$_3$ (shown in Figure-5) with vacancies at various sites as discussed above. The magnified picture of the low energy zone is shown in the figure's inset. Figure 5 shows that there are considerable changes in optical absorption spectra below the basic band gap absorption edge as compared to the pristine material, and that the traces of new states in the optical absorption spectra are easily discernible. Defects at the oxygen site in CaTiO$_3$ are known to produce Ti$^{3+}$ defect centers[35,36]. Littlewood *et al. and* Demkov *et al.* have theoretically predicted that Ti exists in the 3+ state [37–39]. Because of the fundamentally incomplete/broken bonds caused by the oxygen vacancies, weakly confined electrons are causing the system to behave as an electron donor [40]. These donated electrons have been observed to re-hybridize with the Ti orbitals in CaTiO$_3$, converting the Ti$^{4+}$ ions to Ti$^{3+}$ ions [37–41]. To keep the sample's overall charge neutrality, the oxidation-state/charge-state of the transition metal ion must change. Donor levels below conduction bands are known to result from this shift in the oxidation state. [37–39,41]. The similar arguments are true also for Ca and Ti vacancies. The detection of such defect states is crucial from a technological perspective because it is known that these vacancies enhance the electrical and optical capabilities of devices. [42–45]. Hence, signatures of vacancies in semiconducting oxides can be identified by comparing the experimental data with the theoretically computed spectra corresponding to various types of vacancies.

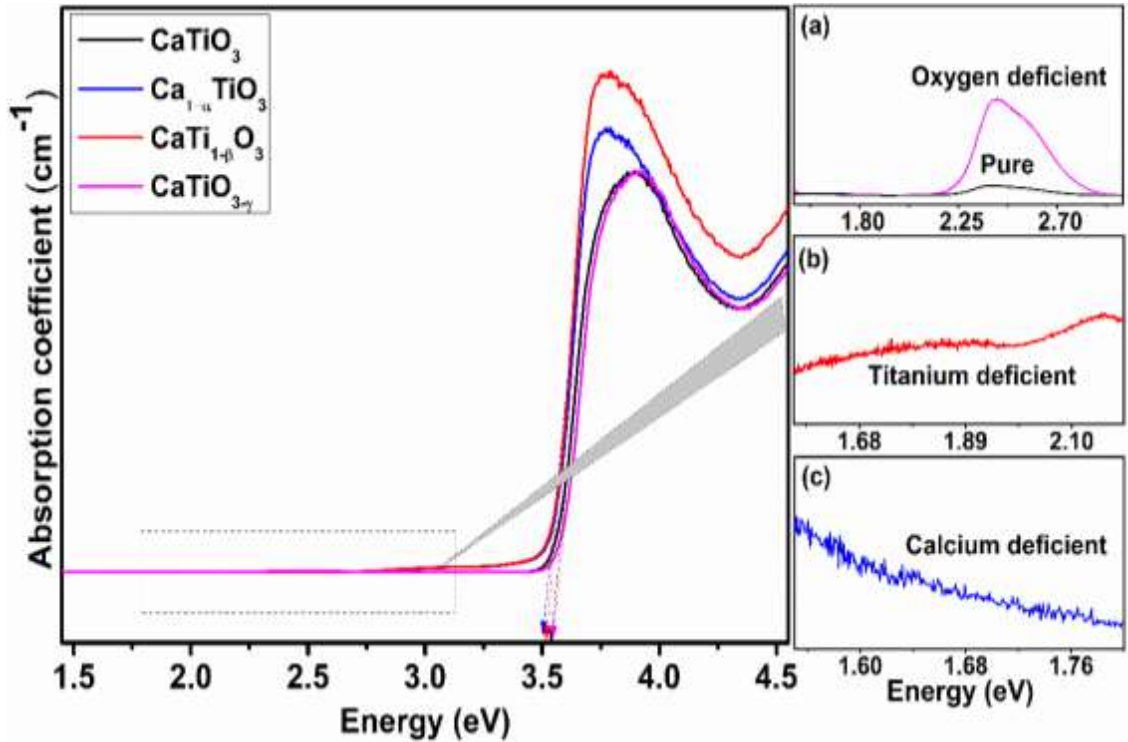

*Figure 5:* Optical absorption spectra of pristine and self- deficient $CaTiO_3$, inset (a) Ti deficient (b) O deficient (c) Ca deficient.

### 3.3. Electrical Measurements:

### 3.3.1 Switching Stability:

In order to probe the stability of our devices, UV photocurrents of pure and self-deficient $CaTiO_3$, at a constant bias voltage of 5 V, were measured under ambient conditions. For the purpose, the photons of wave lengths 280 nm were alternatively switched ''ON'' and ''OFF'' for 500 seconds each, and the results of our measurements are shown in Figure 6. This growth and decay of current can be used to sense UV as well as visible wavelengths. Interestingly, it is found that in case of UV light, photocurrent response shows significantly higher values for the pure and the self-deficient devices, and also shows higher switching stability of all deposited devices for longer time periods.

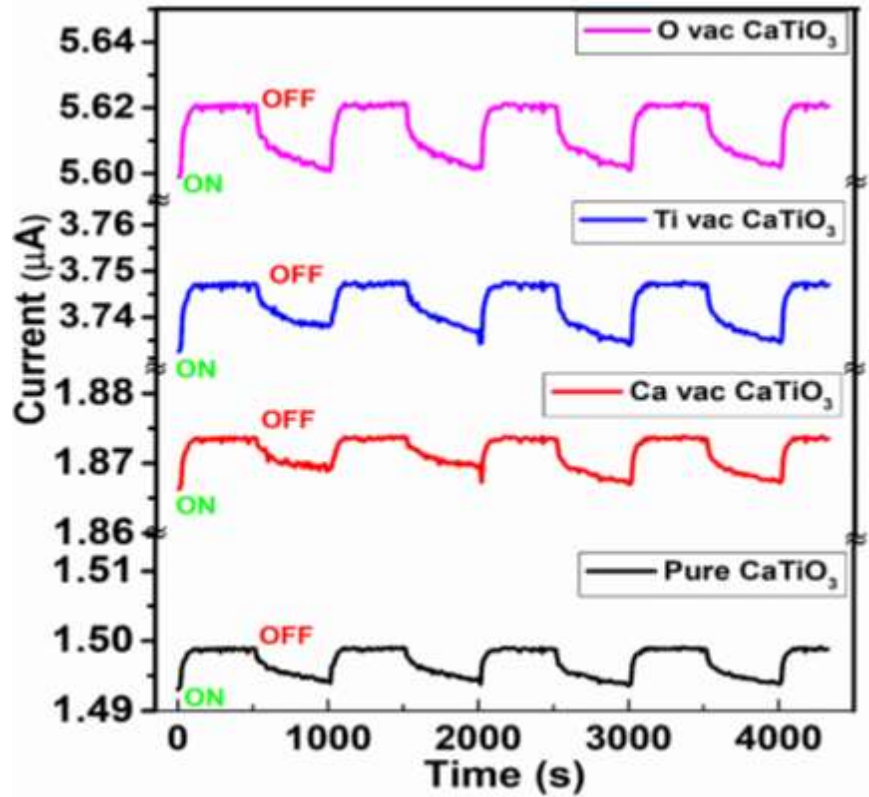

*Figure 6:* *Device stability of pristine and self-deficient CaTiO$_3$.*

**3.3.2. Current Density and Incident-Photon-to-Current Efficiency Measurement:**

Incident-photon-to-current efficiency (IPCE) of prepared device have been measured fir pure and self-deficient CaTiO$_3$ based devices. From Figure 7 (a) it is obvious that IPCE% shows enhancement in the case of vacancy induced calcium titanites as compare to pure CaTiO$_3$. The pure CaTiO$_3$ based device achieved a maximum IPCE of 8-10% while in case of self-deficiency it shows better performance with a higher IPCE of 10-12%, 20-23%, and 25-28% in case of Ca, O, and Ti-deficiency respectively. Hence for the enhancement of efficiency in the semiconducting oxides defective samples are much more important than the pure ones. J-V curves of the prepared pure and self-deficient device at the optimal conditions were depicted in Figure 7 (b). The improved performance can be also reflected in the enhanced J$_{sc}$ values increasing from 30.08 mA/cm$^2$, 38.02 mA/cm$^2$, 41.62 mA/cm$^2$, 48.25 mA/cm$^2$ for pure, Ca, Ti, and O deficient respectively (see Figure 6(b)).

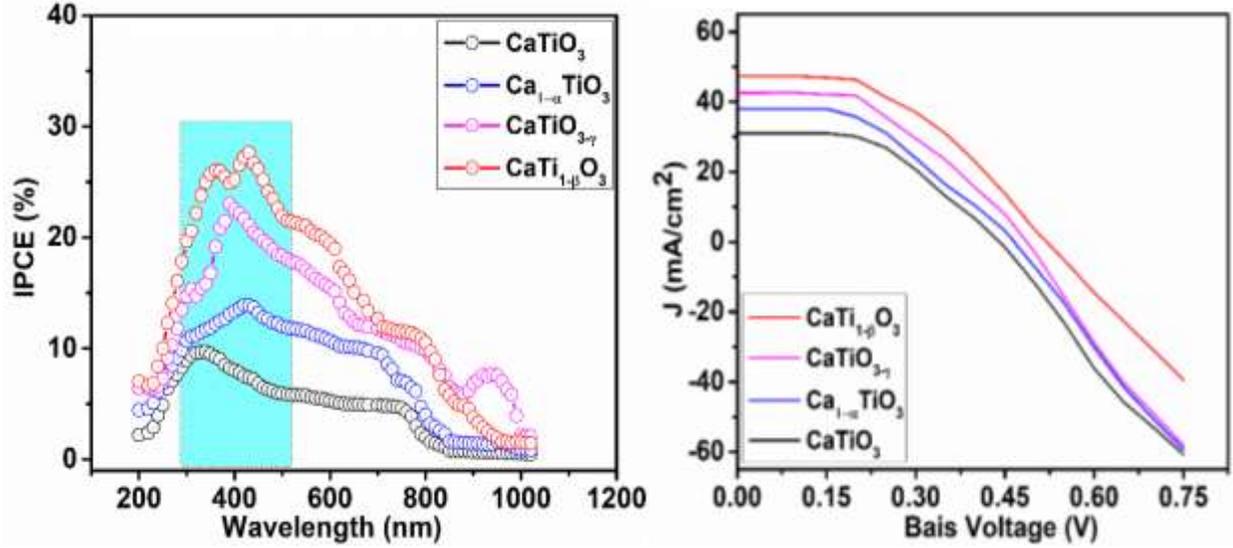

**Figure 7: (a)** *IPCE of pristine and self-deficient CaTiO$_3$* **(b)** *Current density of pristine and self-deficient CaTiO$_3$.*

### 3.3.3. Responsivity and Detectivity Measurement:

To determine the performance of device, responsivity, $R = (I_{UV}/I_P)$ [19], is an important parameter where, $I_P$ is the incident power and $I_{UV}$ is the maximum current under UV irradiation. From Figure 8 (a) it is obvious that responsivity of our devices is maximum in the UV region, while it is decreases in the visible region. The detectivity ($D = 1/NEP_B$) [19] depicted in Figure 8(b) reaches its peak values $98 \times 10^{14}$ Hz$^{1/2}$/W at ~ 280 nm, $18 \times 10^{14}$ Hz$^{1/2}$/W at 260 nm, and $15 \times 10^{14}$ Hz$^{1/2}$/W at 260 nm, for O, Ti, and Ca deficient CTO devices, respectively. Thus, we can conclude that the devices made using self-deficient samples of CaTiO$_3$ show enhancement in both responsivity and detectivity when compared to the ones based on pristine samples.

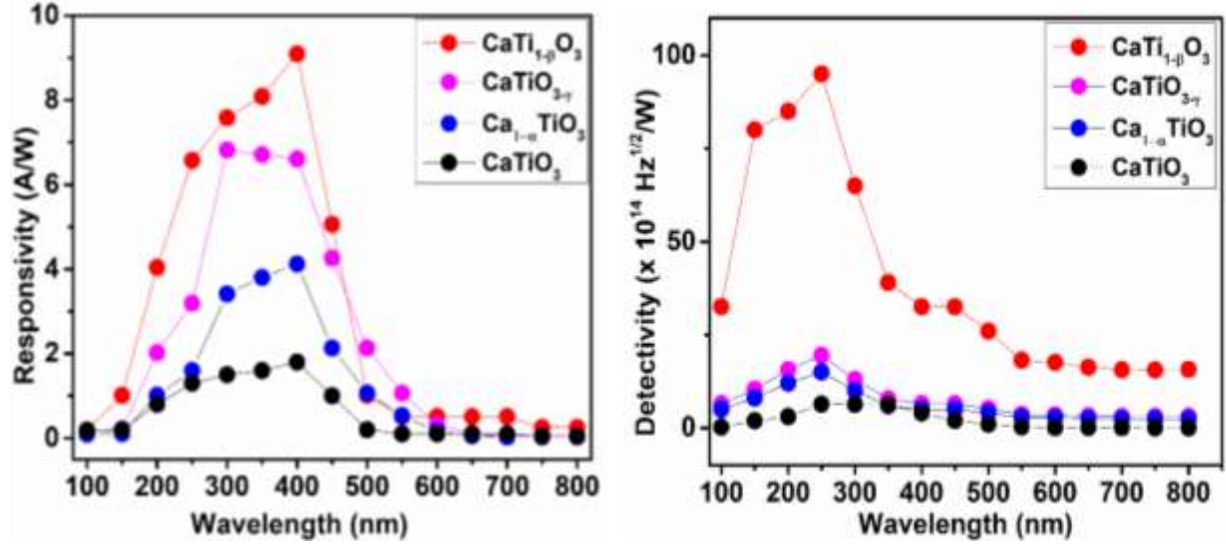

*Figure 8:* (a) *Responsivity and (b) Detectivity of pristine and self-doped CaTiO$_3$.*

### 3.3.4. Electrical Transport and Recombination due to Defect Centers:

The transport resistance ($R_{tr}$) is inversely proportional to carrier conductivity ($\sigma$) while the recombination resistance ($R_{rec}$) depends on the variation of the recombination flux ($J_{rec}$) [46].

$$R_{tr} = L/A\,\sigma \tag{2}$$

$$R_{rec} = \frac{1}{A}\left(\frac{\partial J_{rec}}{\partial V}\right)^{-1} \tag{3}$$

Above A represents active area of film deposited, L denotes layered thickness, while V represents the applied voltage.

The recombination resistance, $R_{rec}$, shown in Figure 9 (a), contains crucial information on recombination in the solar cell. At low voltage, we can see that the self-deficient samples exhibit lower resistance (greater recombination rate) than the pure sample, while at the higher voltages they exhibit almost the same resistance. Figure 9(b) clearly shows higher carrier conductivity for defect-based devices as compared to the pure one, implying that variations in recombination rates and carrier conductivity are playing key roles in the performance differences in the prepared self-deficient devices.

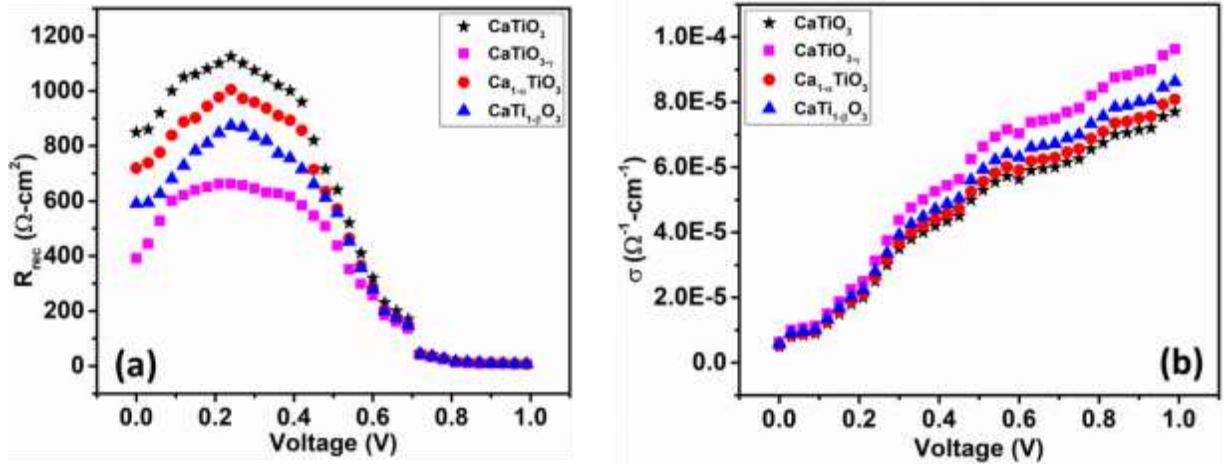

*Figure 9: Transport and recombination parameters vs voltage (a) recombination resistance, $R_{rec}$, (b) conductivity of pure and self-deficient $CaTiO_3$.*

Moreover, a comparison of our devices with the performance metrics of solar cells made from other materials has been summarized in Table 1 [44-46]. From the comparison it is obvious that although CTO based devices demonstrate smaller responsivities, however, they show significantly higher detectivities as compared to those made from other materials.

**Table 1:** *The comparison of the performance of self-deficient $CaTiO_3$-based solar cells, with those made using other materials.*

| Devices | R (A/W) | D ($Hz^{1/2}/W$) | ON/OFF Ratio | $V_{oc}$ | FF | Ref. |
|---|---|---|---|---|---|---|
| $BaTiO_3$ hetrostructures | 20 | $3 \times 10^9$ | 5.2 | 1.05 | 0.71 | [47] |
| ZnO thin films | 60.5 | $1.48 \times 10^{14}$ | - | 1.02 | 0.70 | [48] |
| $MAPbI_3$ composite films | ~30 | $2.4 \times 10^{14}$ | - | 1.06 | 0.76 | [49] |
| $CaTiO_3$ | 1.32 | $12 \times 10^{14}$ | $2.61 \times 10^2$ | 1.01 | 0.62 | |
| $Ca_{1-\alpha}TiO_3$ | 1.5 | $15 \times 10^{14}$ | $2.75 \times 10^2$ | 1.05 | 0.75 | This Work |
| $CaTi_{1-\beta}O_3$ | 6.5 | $18 \times 10^{14}$ | $3.10 \times 10^2$ | 1.08 | 0.78 | |
| $CaTiO_{3-\gamma}$ | 9 | $98 \times 10^{14}$ | $3.46 \times 10^2$ | 1.09 | 0.79 | |

*MA = methyl amine ($CH_3NH_3$-)

## 4. Conclusion:

In summary we have shown that vacancies at Ca, Ti and O site in $CaTiO_3$ lead to the extra states in optical absorption spectra. Enhancement in IPCE and current density has been observed for all deficient $CaTiO_3$ devices as compared to the pristine ones. In samples with Ti and O vacancies we observed maximum IPCE in the range 25%-28 % (280-400 nm) in the UV region. It has been determined how self-deficiency affects responsivity and detectivity of a solar cell. By combining experimental and theoretical investigations it is evident that self-deficiency has very high impact on the opto-electronic properties of solar cells based on perovskites. Therefore, our observation that the perovskite-based solar cells with native defects perform with better stability and large efficiency will lead to important advances in the semiconductor industry.

**Acknowledgments**

The Homi Bhabha Research Cum Teaching Fellowship (A.K.T.U.), Lucknow, India, is acknowledged by one of the authors (SP) for providing financial support in the form of a teaching assistantship. The authors are extremely grateful to Prof. M. S. Ramachandra Rao (IIT Madras) and Dr. Tejendra Dixit (IIIT Kancheepuram) for their assistance with the synthesis and for providing the experimental facilities.

**Data availability statement –**

The raw/processed data required to reproduce these findings cannot be shared at this time as the data also forms part of an ongoing study.

**Compliance with ethical standards:**

Conflict of interest: The authors declare that they do not have any conflict of interest.